\documentclass[oribibl]{llncs2015}
\usepackage{filecontents}

\usepackage[
  style=alphabetic,
  abbreviate=true,
  firstinits=true,
  url=false,
  isbn=false
  ]{biblatex}

\AtEveryBibitem{%
  \clearname{translator}%
  \clearfield{pagetotal}%
  \clearfield{booktitle}%
  \clearfield{issn}%
  \clearfield{groups}%
  \clearfield{journaltitle}
  \clearfield{volume}
  \clearfield{number}
  \clearname{editor}
}
\DeclareFieldFormat
  [article,inbook,incollection,inproceedings,patent,thesis,unpublished]
  {title}{#1}

\usepackage{float}
\usepackage{hyperref}
\usepackage{graphicx}
\usepackage{amsmath}
\usepackage{amsfonts}
\usepackage{latexml}
\usepackage{color}
\newcommand{\theReferenceText}{\fullcite{SchubotzMHCG17}}
\usepackage{comgipp}
\usepackage[subtle,tracking=normal]{savetrees}
\definecolor{gray}{rgb}{0.4,0.4,0.4}
\definecolor{darkblue}{rgb}{0.0,0.0,0.6}
\definecolor{cyan}{rgb}{0.0,0.6,0.6}
\definecolor{darkgreen}{rgb}{0,0.5,0}

\usepackage{listings}
\usepackage{lstlinebgrd}

\lstdefinelanguage{XML}
{
  morecomment=[f][\color{red}]{-\ },         
  morestring=[b]",
  morestring=[s]{>}{<},
  morecomment=[s]{<?}{?>},
  morecomment=[s]{!--}{--},
  morecomment=[s][\color{red}]{\$}{\$},
  commentstyle=\color{darkgreen},
  stringstyle=\color{black},
  identifierstyle=\color{darkblue},
  keywordstyle=\color{cyan},
  basicstyle=\footnotesize,
  morecomment=[f][\color{green}]+,       
  morekeywords={xmlns,version,type}
}

\begin{document}

\title{VMEXT: A Visualization Tool for Mathematical Expression 
Trees}
\titlerunning{VMEXT}

\author{Moritz Schubotz\inst{1} \and Norman Meuschke\inst{1} \and Thomas Hepp\inst{1} \and
   \\Howard S. Cohl\inst{2} \and Bela Gipp\inst{1}}
\authorrunning{Schubotz et al.}

\institute{Dept. of Computer and Information Science,\\University of Konstanz,
  \\Box 76, 78457 Konstanz, Germany\\
  \email{\{first.last\}@uni-konstanz.de}, \texttt{www.isg.uni-konstanz.de}
  \vspace{0.25cm}
  \and
  Applied and Computational Mathematics Division,\\National Institute of Standards and Technology\\
  Gaithersburg, Maryland 20899-8910, USA\\
  \email{howard.cohl@nist.gov}, \texttt{www.nist.gov/people/howard-cohl}
}
\maketitle
\thispagestyle{firststyle}
\begin{abstract}
Mathematical expressions can be represented as a tree consisting of terminal symbols, such as
identifiers or numbers (leaf nodes), and functions or operators (non-leaf nodes). Expression trees
are an important mechanism for storing and processing mathematical expressions as well as the most
frequently used visualization of the structure of mathematical expressions. Typically, researchers
and practitioners manually visualize expression trees using general-purpose tools. This approach is
laborious, redundant, and error-prone. Manual visualizations represents a user's notion of what the
markup of an expression should be, but not necessarily what the actual markup is. This paper
presents VMEXT -- a free and open source tool to directly visualize expression trees from parallel
\MathML{}. VMEXT simultaneously visualizes the presentation elements and the semantic structure of
mathematical expressions to enable users to quickly spot deficiencies in the Content \MathML{}
markup that does not affect the presentation of the expression. Identifying such discrepancies
previously required reading the verbose and complex \MathML{} markup. VMEXT also allows one to
visualize similar and identical elements of two expressions. Visualizing expression similarity can
support support developers in designing retrieval approaches and enable improved interaction
concepts for users of mathematical information retrieval systems. We demonstrate VMEXT's
visualizations in two web-based applications. The first application presents the visualizations
alone. The second application shows a possible integration of the visualizations in systems for
mathematical knowledge management and mathematical information retrieval. The application converts
\LaTeX\ input to parallel \MathML{}, computes basic similarity measures for mathematical
expressions, and visualizes the results using VMEXT.
\end{abstract}

\section{Introduction}\label{sec:intro}
Mathematical notation strives to have a well-defined vocabulary, syntax, and semantics. Similar to
sentences in natural language or constructs in a programming language, mathematical expressions
consist of constituents that have a coherent meaning, such as terms or functions. We consider a
mathematical expression to be any sequence of mathematical symbols that can be evaluated, e.g.,
typically formulae. The syntactic rules of mathematical notation, such as operator precedence and
function scope, determine a hierarchical structure for mathematical expressions, which can be
understood, represented, and processed as a tree. \textit{Mathematical expression trees} consist of
functions or operators and their arguments. Experiments by \citeauthor{jansen2000} suggest that
mathematicians use some notion of mathematical expression trees as a mental representation to
perform mathematical tasks \autocite{jansen2000}.

Describing and processing mathematical content using expression trees is established practice in
mathematics and computer science as our review of related work in Section \ref{sec:rel} shows.
However, no standard for the content of nodes, or the structure and visual representation of such
trees has yet emerged. Additionally, we did not find tools that support generating expression tree
visualizations from mathematical markup. All of the visualizations that we were able to glean from
the literature were manually created using general purpose tools.

With this paper, we seek to contribute to the establishment of an openly available, widely accepted,
visualization of mathematical expression trees, encoded using the \MathML{} standard. For this
purpose, we propose a tree visualization that operates on parallel \MathML{} markup and provides the
visualization as a free and open source tool. We structure the presentation of our contributions as
follows. Section \ref{sec:rel:mathml} presents details of the \MathML{} standard that serves as the
data structure for our visualization approach. Section \ref{sec:rel:vis} reviews the strength and
weaknesses of existing approaches for visualizing mathematical expression trees to derive our
visualization concept. Section \ref{sec:vmext} present our visualization tool VMEXT. Section
\ref{sec:vmext:demo} describes a demo application that shows how the visualization can be integrated
into other applications. Section \ref{sec:vmext:use} explains how end users and developers can apply
and obtain VMEXT. Section \ref{sec:future-work} concludes the paper by discussing our plans for
further extending and improving VMEXT.

\section{Related Work}\label{sec:rel}
As briefly motivated in the previous section, we seek to reduce the effort for researchers and
practitioners to generate expression tree visualizations for mathematical content. Additionally, we
hope to contribute to establishing a standardized representation of mathematical expression trees.
In Section \ref{sec:rel:mathml}, we present the \MathML{} standard and explain why we see it as a
promising data format to achieve this goal. In Section \ref{sec:rel:vis}, we review existing
approaches for visualizing mathematical expression trees to explain how we derived the major
building blocks of our visualization approach.

\subsection{\MathML{}}\label{sec:rel:mathml}
Mathematical Markup Language (\MathML{}) is a W3C\footnote{\url{www.w3.org/Math/}} and ISO standard
(ISO/IEC DIS 40314) for representing mathematical content using XML syntax. \MathML{} is part of
HTML5 and enables one to serve, receive, and process mathematical content on the World Wide Web.
\MathML{} allows users to describe the notation and/or the meaning of mathematical content using two
vocabularies: Presentation \MathML{} (PMML) and Content \MathML{} (CMML). The vocabularies can be
used independently of each other or in conjunction.

Presentation \MathML{} focuses on describing the visual layout of mathematical content. The PMML
vocabulary contains elements for basic mathematical symbols and structures. Each element specifies
the role of the presentation element, e.g., the element \texttt{\textless mi\textgreater} represents
identifiers and the element \texttt{\textless mo\textgreater} represents operators. The structure of
PMML markup reflects the two-dimensional layout of the mathematical expression. Elements that form
semantic units are encapsulated in \texttt{\textless mrow\textgreater} elements, which are
comparable to \texttt{\textless div\textgreater} elements in HTML. \autoref{lst:pmml} exemplifies
PMML markup for the expression $f(a+b)$.

Content \MathML{} focuses on explicitly encoding the semantic structure and the meaning of
mathematical content using expression trees. In other words, the CMML vocabulary seeks to specify
the frequently ambiguous mapping from the presentation of mathematical content to its meaning. For
example, the presentation of the expression $f(a+b)$ represents two possible syntactic structures:
e.g., $f$ could represent either an identifier or a function. CMML uses \texttt{\textless
  apply\textgreater} elements to make explicit which elements represent functions. Subordinate
elements represent the arguments of the functions. \autoref{lst:cmml} illustrates CMML markup for
the expression $f(a+b)$.

\begin{lstlisting}[language=XML,label=lst:pmml,caption={Presentation \MathML{} encoding of the expression $f(a+b)$ \cite{dis}}]
<math xmlns="http://www.w3.org/1998/Math/MathML">
  <semantics>
    <mrow id="r1">
      <mi id="i1">f</mi>
      <mo id="o1">(</mo>
      <mrow id="r2">
        <mi id="i2">a</mi>
        <mo id="o2">+</mo>
        <mi id="i3">b</mi>
      </mrow>
      <mo id="o3">)</mi>
    </mrow>
\end{lstlisting}
Content \MathML{} offers two subsets of elements to specify function types: Pragmatic Content
\MathML{} and Strict Content \MathML{}. Pragmatic Content \MathML{} uses a large set of predefined
functions encoded as empty elements, e.g., \texttt{\textless plus/\textgreater}, as used in Line 17
in \autoref{lst:cmml}, or \texttt{\textless log/\textgreater} for the logarithm. Strict Content
\MathML{} uses a minimal set of elements, which are further specified by referencing extensible
content dictionaries. For example, the plus operator (+) is defined in the content dictionary
\texttt{arith1}. Using Strict CMML, the operator is encoded using the element for symbols
\texttt{\textless csymbol\textgreater}, and declaring that the specification of the symbol is
available under the term \texttt{plus} in the content dictionary \texttt{arith1}. Line 17 in
\autoref{lst:cmml} shows this option of specifying the plus operator as a comment (green font
color).

\begin{lstlisting}[float,floatplacement=h,language=XML,label=lst:cmml,firstnumber=13,
  caption={Content \MathML{} encoding of the expression $f(a+b)$ \cite{dis}}]
    <annotation-xml encoding="MathML-Content">
      <apply xref="r1">
        <ci xref="b">f</ci>
        <apply xref="r2">
          <plus xref="o2"/><!-- <csymbol cd="arith1">plus</csymbol> in strict encoding -->
          <ci xref="i2">a</ci>
          <ci xref="i3">b</ci>
        </apply>
      </apply>
    </annotation-xml>
\end{lstlisting}
As described above, the PMML and CMML vocabularies can be used individually and independent of each
other. For example, PMML is frequently used without content markup to display mathematical content
on websites. CMML without presentation markup can, for instance, be used to exchange data between
computer algebra systems. However, PMML and CMML markup can also be used in conjunction to
simultaneously describe the presentation, structure, and semantics of mathematical expressions. The
combined use of PMML and CMML is commonly referred to as parallel \MathML.

In parallel \MathML{} markup, presentation and content elements are mutually interlinked by
including \texttt{xref} arguments that point to the corresponding element in the other vocabulary.
The PMML and CMML markup in \autoref{lst:pmml} and \autoref{lst:cmml} respectively contain
\texttt{xref}-links to create parallel \MathML{}.

\subsection{Expression Tree Visualizations}\label{sec:rel:vis}
\begin{figure}[!ht]
        \begin{center}
        \caption{Overview of expression tree visualizations part 1}
                \includegraphics[width=\linewidth]{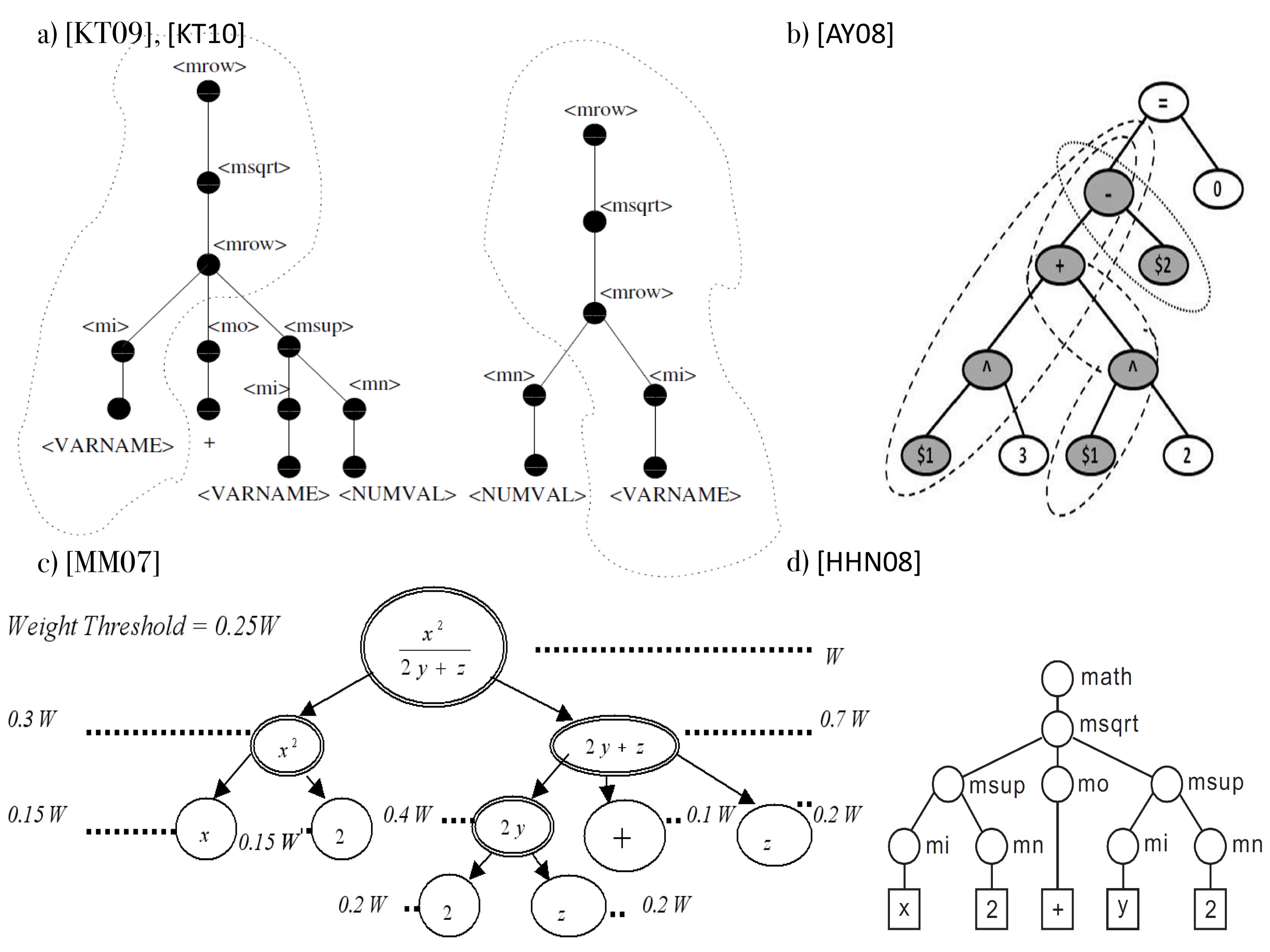}
                \label{fig:collage1}
        \end{center}
\end{figure}

\begin{figure}[h]
        \begin{center}
        \caption{Overview of expression tree visualizations part 2}
                \includegraphics[width=\linewidth]{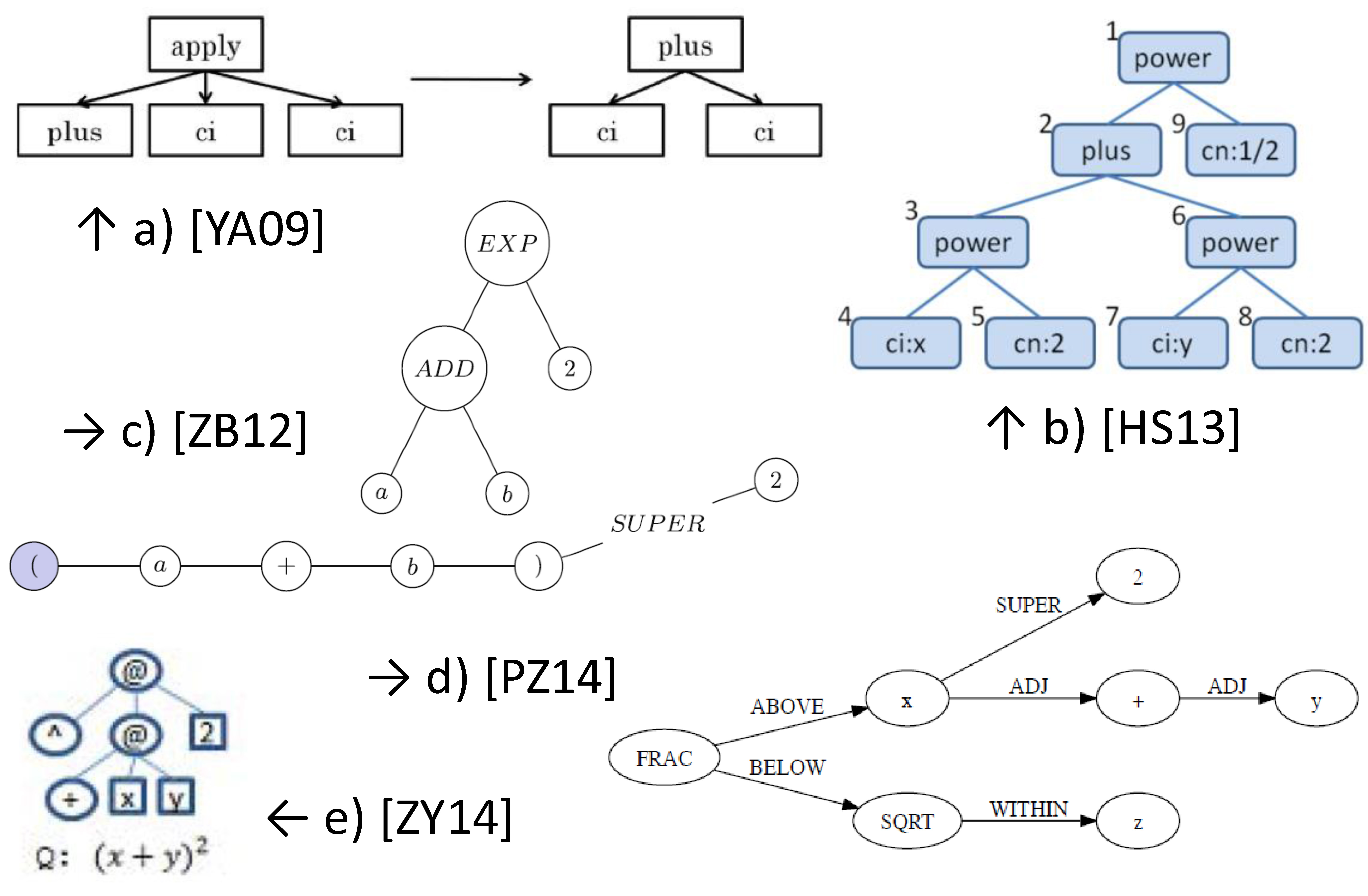}
                \label{fig:collage2}
        \end{center}
\end{figure}

Researchers, especially in math information retrieval (MIR), have employed several use-case-specific
tree visualizations for mathematical expressions. All visualizations appear to have been created
manually to illustrate research in publications. The content and structure of the visualizations
vary significantly. \autoref{fig:collage1} and \autoref{fig:collage2} give an overview of the
visualizations, which we describe hereafter.

\citeauthor{youssef2006math} use simple ASCII graphics to visualize expression trees. Their
visualization resembles binary expression trees. Leaf nodes represent identifiers or numbers; inner
nodes represent operators, functions, or brackets \autocite{youssef2006math}. In later work,
\citeauthor{shatnawiY07} replace the ASCII graphics with an equivalent chart.
\citeauthor{altamimiY08} further improve their visualization by marking subexpression groups with
dashed lines (see \autoref{fig:collage1}, b) \autocite{altamimiY08}.

\citeauthor{minerM07} use a different tree to illustrate their research on math search. They render
the full expression in the root of the tree and create sub-nodes for each sub-expression (see
\autoref{fig:collage1}, c) \autocite{minerM07}. \citeauthor{sojka2011art} use a similar
visualization to illustrate the tokenization and indexing process of their math search system.

\citeauthor{hashimotoHN08} use a tree layout that represents the DOM structure of Presentation
\MathML{} markup to illustrate the author's research on \MathML{} indexing \autocite{hashimotoHN08}.
In this layout, inner nodes represent \MathML{} elements depicted as circles and leaf nodes
represent the content of elements depicted as squares (see \autoref{fig:collage1}, d). We assume the
authors manually created the visualization, since the focus of their paper is on math search and
does not mention an automated visualization approach.

\citeauthor{kamali2009improving} \autocite{kamali2009improving} and \citeauthor{kamaliT10}
\autocite{kamaliT10} use a similar tree representation of the Presentation \MathML{} structure in
their works on math similarity and retrieval. Their visualization does not distinguish between inner
nodes and leaf nodes, but depicts all nodes as circles (see \autoref{fig:collage1}, a). Two things
are notable about this visualization. First, the layout corresponds to the data structure of the
mathematical expressions. Second, \citeauthor{kamali2009improving} introduce the notion of defining
and visualizing the similarity of mathematical expressions in terms of the structural similarity of
sub-trees. The authors visually indicate similar sub-trees by enclosing the respective sub-tree in a
dashed line (see \autoref{fig:collage1}, a). In subsequent work, \citeauthor{kamaliT13}
\autocite{kamaliT13} use a horizontal layout to visualize the same tree. The tree uses boxes instead
of circles and directed instead of undirected edges. \citeauthor{kamaliT13} exclusively consider
PMML and do not present an automated approach to create their visualization of the structure and
similarity of PMML expressions.

\citeauthor{yokoi2009} consider Content \MathML{} markup for their research on math similarity
search and devise a visualization of the CMML tree \autocite{yokoi2009}. Their work introduces
apply-free content markup, i.e., omitting the first \texttt{\textless apply\textgreater} element in
the CMML markup, since it provides little information on the applied function. Instead, their markup
uses the first child of an \texttt{\textless apply\textgreater} element. Their manually created
visualization also omits \texttt{\textless apply\textgreater} elements (see \autoref{fig:collage2},
a). We consider this approach valuable, since it reduces the number of nodes to visualize and
facilitates the recognition of function types.

\citeauthor{haginoS13} also consider apply-free Content \MathML{} markup for their research on
partial match retrieval in math search \autocite{haginoS13}. To illustrate their research, they use
a tree that depicts the CMML element names in the case of inner nodes and the CMML element names in
combination with the elements' content in the case of leaf nodes (see \autoref{fig:collage2}, b).

In their review of approaches for math recognition and retrieval, \citeauthor{zanibbiB12} point out
that building a symbol layout tree is important for math recognition tasks \autocite{zanibbiB12}.
Symbol layout trees represent horizontally adjacent symbols that share a writing line and indicate
subscript, superscript, above, below, and containment relationships. The authors present a
horizontal illustration of the symbol layout tree and a simplified expression tree using a vertical
layout (see \autoref{fig:collage2}, d). \citeauthor{pattaniyilZ14} uses a similar horizontal
illustration of the symbol layout tree (see \autoref{fig:collage2}, e) \autocite{pattaniyilZ14}.

\citeauthor{zhangY14} use Strict Content \MathML{} for their research \autocite{zhangY14}. In their
visualizations of the CMML tree, they omit the element names for \texttt{\textless ci\textgreater}
and \texttt{\textless cn\textgreater} elements, but include \texttt{\textless apply\textgreater}
elements. They replace the names of CMML elements with shorter symbols. For instance, they replace
\texttt{\textless apply\textgreater} with \texttt{@} and \texttt{\textless power\textgreater} with
\texttt{$^\wedge$}.

\subsection{Summary of Related Work and Research Gap}
From our review of the literature, we draw the following conclusions. First, representing
mathematical expressions as trees is essential for performing many tasks in mathematical knowledge
management (MKM) and mathematical information retrieval (MIR). Expression trees, in which leaf nodes
represent terminal symbols and inner nodes represent operators, functions, or brackets are widely
used as a data representation. The \MathML{} standard is a well-established data format for
representing the presentation, structure, and semantics of mathematical content using the expression
tree concept. Many researcher rely on \MathML{} encoded content for MIR and MKM tasks.

Second, researchers frequently employ expression tree visualizations to illustrate their
math-related research. While some visualizations reflect the information extracted from mathematical
markup, such as \MathML, other visualizations illustrate abstract mathematical expressions. The
elements included in the visualizations, their spatial arrangement, and visual appearance varies
greatly. Depending on the use case, visualizations may include presentation elements, content
elements, or combinations thereof. Especially in the MIR domain, researchers frequently need to
visualize similarity of operator (sub-)trees.

Third, although the expression tree concept is at the heart of \MathML{} and visualizations of
\MathML{} markup are widely used for analysis and presentation purposes, we found no tool that
generates such visualizations from \MathML{} markup. Researchers typically create expression tree
visualizations manually using general-purpose tools. This approach results in much manual and
redundant effort, diverse visual representations of identical markup, and the danger of creating a
visualization that does not reflect the underlying data. To reduce the effort for creating
expression tree visualizations and to contribute towards establishing a more canonical design of
expression trees, we present the VMEXT system, which we describe in the following section.

\section{VMEXT System}\label{sec:vmext}
VMEXT is an acronym for Visualizing Mathematical Expression Trees. This tool seeks to visually
support researchers and practitioners in two well-defined use cases:

\begin{enumerate}
  \item curating semantically enriched mathematical content, e.g., for use in digital repositories
    or systems for mathematical knowledge management;
  \item examining similarities of two mathematical expressions, e.g., for developing mathematical
    information retrieval approaches or for examining and interacting with the results of MIR
    systems.
\end{enumerate}

VMEXT addresses the use cases with two visualizations available as widgets that can easily be
integrated into any web application. We present the widgets in Section \ref{sec:vmext:ast} and
Section \ref{sec:vmext:sim}. Both widgets are available as a demo system at:
\url{http://vmext.formulasearchengine.com/}. Section \ref{sec:vmext:demo} presents a demo
application that exemplifies the possible use of the widgets as part of MKM and MIR systems. Section
\ref{sec:vmext:use} describes how interested parties may use VMEXT's visualizations; integrate the
visualizations as widgets or via an API into their own applications; and how to adapt and extend the
code.

\subsection{Curating Semantically-enriched Mathematical Content}\label{sec:vmext:ast}
Making mathematical knowledge accessible through recognition, retrieval, and management systems is a
task that has attracted many contributions by researchers and practitioners. (\citeauthor{guidi2016}
\autocite{guidi2016} and \citeauthor{zanibbiB12} \autocite{zanibbiB12} present comprehensive reviews
on the topic). The \MathML{} standard (see Section \ref{sec:rel:mathml}) has been widely adopted to
expose both the presentation and semantics of mathematical content for such systems.

However, the \MathML{} syntax is verbose, complex and therefore not easy to grasp for humans.
Furthermore, creating parallel \MathML{} markup is complicated and error-prone. This is true,
especially for the creation of parallel \MathML{} by converting other formats, such as \LaTeX, and
often results in ambiguous or erroneous markup. Typically, Presentation \MathML{} elements are less
frequently affected by errors than their respective Content \MathML{} elements. This leads to a
situation, in which the visual representation of an expression is correct, yet its semantics are
wrong.

VMEXT supports users in quickly checking and improving parallel \MathML{} by providing an 
interactive expression tree visualization that simultaneously illustrates the semantic structure 
(as well as the presentation elements) encoded in the markup.

\begin{figure}[h]
\includegraphics[width=\textwidth]{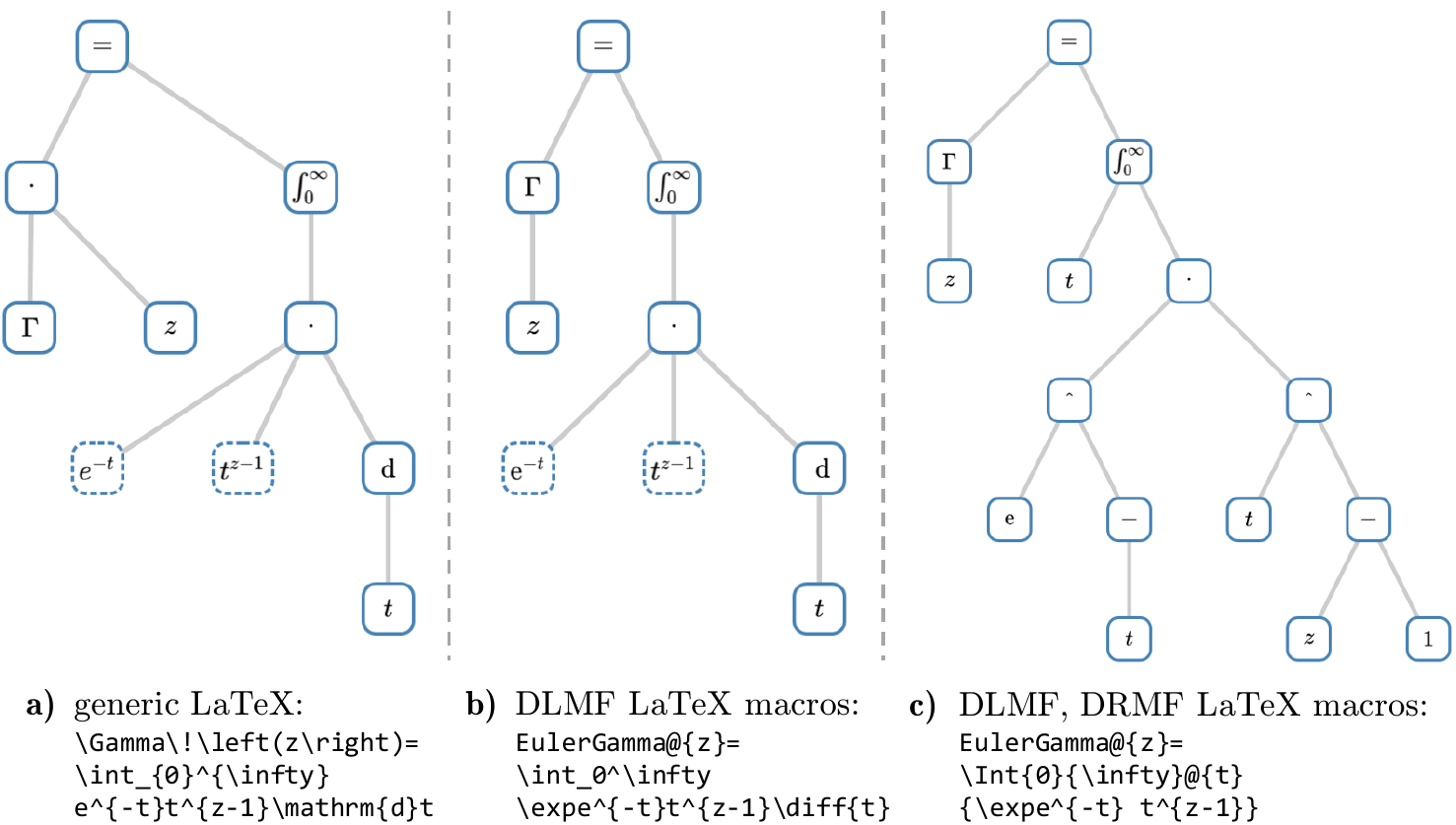}
\caption{Expression trees rendered for \MathML\ input obtained from converting different \LaTeX\
  input. The Presentation \MathML\ is identical for all three cases, yet the Content \MathML\
  differs.}
\label{fig:euler}
\end{figure}

VMEXT visualizes the structure of the tree as encoded in the Content \MathML{} markup. However, the
labels for each node render the Presentation \MathML{} elements linked to the respective content
elements. VMEXT uses the apply-free CMML notation introduced in \autocite{yokoi2009}. In other
words, our parser renders the first child of each \texttt{\textless apply\textgreater} element, not
the \texttt{\textless apply\textgreater} itself, as an operator or function. All following children
are considered as arguments of the function. For a clear layout, VMEXT renders the complete PMML
element for the first child, even if the first child is itself an \texttt{\textless
  apply\textgreater} element. To reduce the size of the individual edges, we replace some CMML
elements with shorthand symbols, e.g., we replace \texttt{\textless power\textgreater} with $\wedge$
as can be seen in \autoref{fig:euler} (cf. \autocite{zhangY14}, see also Section \ref{sec:rel}).

To facilitate human inspection, VMEXT follows the information seeking mantra proposed by
\citeauthor{shneiderman1996} \autocite{shneiderman1996}: \textit{overview first}, \textit{zoom and
filter}, then \textit{details-on-demand}. The nodes in VMEXT can be interactively \textit{filtered}
by expanding or collapsing nodes either one at a time or all at once using the expand button. The
view-port is adjustable using \textit{pan} and \textit{zoom} interactions to enable focusing on
specific parts of the tree. The resize button resets the zoom level. User \textit{navigation} is
supported through an overview infix expression rendered at the top of the screen. Hovering over
parts of the infix expression or nodes in the tree, highlights the corresponding parts in the tree
and the infix expression. \autoref{fig:sim_widget} shows how hovering over the divide operator
highlights the respective sub-tree in light blue. The user can \textit{export} the chosen (sub-)tree
rendering, including all manipulations performed through filtering and zooming, as a high-resolution
{\tt png} image, e.g., for use in publications.

To demonstrate how VMEXT's expression tree visualization can aid in curating semantically enriched
\MathML\ markup, we use the integral representation of the Euler gamma function \cite[(5.2.1)]{dlmf}
as an example

\begin{align}\label{eq:euler}
\Gamma(z)=\int_{0}^{\infty}e^{-t}t^{z-1}\,\mathrm{d}t.
\end{align}
Figures \ref{fig:euler} a-c show VMEXT's rendering for three markup variants of the Euler gamma
function. All variants have identical PMML markup, i.e., produce identical visual output as shown in
\autoref{eq:euler}. However, the CMML differs, because we generated the \MathML\ using \LaTeXML\
\autocite{Miller15} using different \LaTeX\ input (shown in the captions of the figures). Note, that
these different \LaTeX\ versions encode more or less semantics.

The trees in Figures \ref{fig:euler} a and b show that VMEXT allows an arbitrary number of child 
nodes,
as opposed to the binary expression tree concept we briefly described in Section \ref{sec:intro}.
The conversion of generic \LaTeX{} input (a), misinterpreted some invisible operators, such as the
invisible operator between $\Gamma$ and $\left( z \right)$ that was interpreted as times rather than
a function application. Additionally, \LaTeXML\ marked some CMML elements as ambiguous, i.e., could
not establish a one-to-one relation to a PMML element. For ambiguous nodes, VMEXT renders all PMML
elements enclosed by the ambiguous CMML element in a node with dashed borders to emphasize the
defective markup for the user. For example, the node for $e^{-t}$ in \autoref{fig:euler} was marked
as ambiguous.

The \LaTeX\ representation using DLMF macros (b) resolves the problem of invisible operators by
using the @ symbol to make such operators explicit. However, this representation still results in
ambiguous nodes. Representing the Euler gamma function using DLMF and DRMF macros
\cite{disCicm14Drmf, disCicm15} results in correct CMML markup. In (c), we specify the integral
using the semantic macro \texttt{\textbackslash Int} rather than the generic \texttt{\textbackslash
int} command. We have required that all occurrences of the $\wedge$-operator must denote the power
operator. Note that, in order to make this workable, one must create beneficial custom semantic
macros for all other uses of the $\wedge$-operator. These include matrix operations ($A^\dag$),
labeling ($x^\ast$), function spaces ($C^k$), norms ($L^p$), sums ($\sum_{n=0}^\infty$), products
($\prod_{n=0}^\infty$), derivatives ($f^{(2)}(x)$), etc.

By rendering the expression tree as encoded by the CMML markup, VMEXT enables users to quickly spot
markup deficiencies and illuminates the effects of using different conversions or manually changing
markup.

\subsection{Examining Similarities of Mathematical Expressions}\label{sec:vmext:sim}
\begin{figure}[t]
  \begin{center}
    \includegraphics[width=0.7\linewidth]{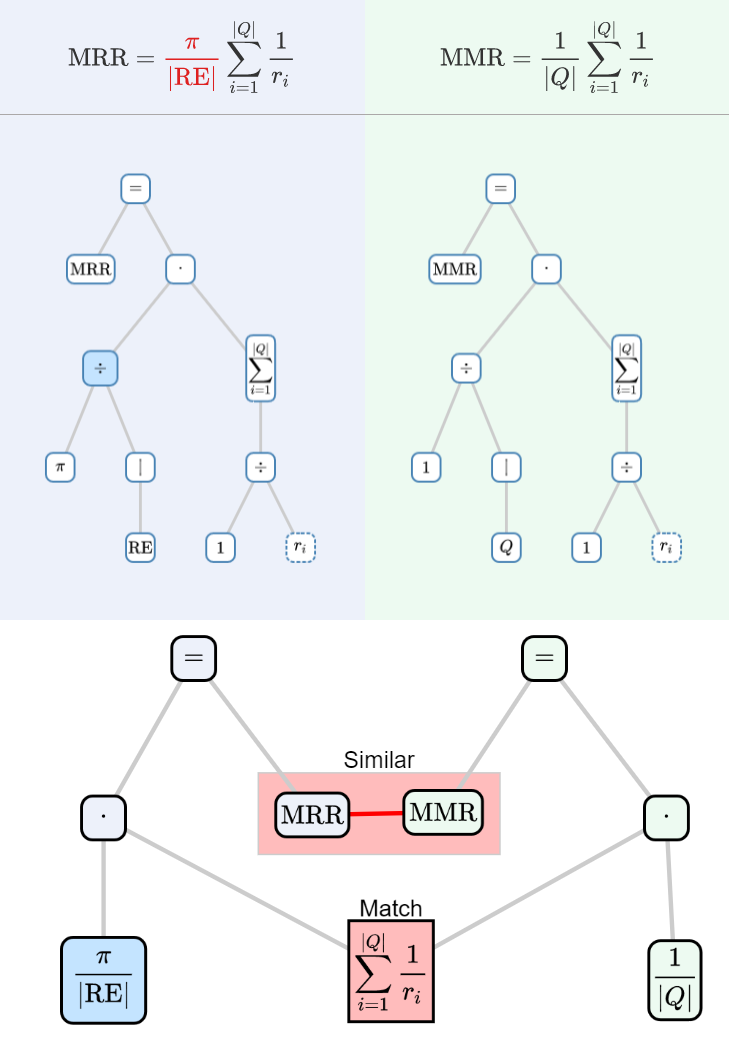}
    \label{fig:sim_widget}
    \caption{VMEXT Expression Tree Similarity Widget}
  \end{center}
\end{figure}

Our review of MIR literature (see Section \ref{sec:rel:vis}) shows that researchers often seek to
visualize the similarity of two mathematical expressions, e.g., the similarity between a query
expression and a retrieval candidate. To facilitate this task, VMEXT includes a specialized
visualization shown in \autoref{fig:sim_widget}. The presented example compares two notations of the
measure Mean Reciprocal Rank.

The widget accepts CMML input for the expressions to compare. Similar elements can be specified by
stating the IDs of the similar CMML elements in both trees using JSON. Currently, VMEXT allows one
to specify that elements are either similar or identical. The two types of similarity are rendered
differently. Since VMEXT is designed to be a visualization tool, it includes no functionality to
compute similarities. We demonstrate the integration of the widgets with a basic application that
computes similarities in Section \ref{sec:vmext:demo}.

The center view renders the trees (including the infix overview) for both expressions and visually
distinguishes the trees using different background colors. The visualizations offer the same
interaction features as the expression tree widget (see Section \ref{sec:vmext:ast}). In the lower
part of the center view, VMEXT renders a combined expression tree. The combined tree includes all
nodes from both trees color-coded with the background color of the tree from which they originate.
Unique, i.e., dissimilar, sub-trees of both trees are collapsed to direct the user's attention to
the similar parts of the trees. For elements marked as similar, VMEXT renders the nodes from both
trees and highlights them as exemplified by the nodes MRR and MMR. Nodes that are marked as
identical are rendered only once and are highlighted as exemplified by the node
$\sum_{i=1}^{|Q|}\frac{1}{r}$.

The integrated visualization of the two expression trees and the combined tree, allows users to
quickly inspect the full structure of both expressions and similar sub-trees. The highlight on hover
feature helps users to look up the corresponding subtrees for nodes marked as similar in the
combined tree.

A specific application that benefits from visualizing the similarity of mathematical expressions is
our prototype of a hybrid plagiarism detection system
CitePlag\footnote{\url{http://www.citeplag.org}} \cite{Meuschke12,Gipp13}. Forms of academic
plagiarism vary greatly in their degree of obfuscation ranging from blatant copying to strongly
disguised idea plagiarism \cite{Meuschke13}. Our research indicates that not a single, but combined
PD approaches are most promising to reliably detect the wide range of plagiarism forms
\cite{Gipp13b,Gipp14,Gipp14a}. Combined approaches accumulate evidence on potentially suspicious
similarity using heterogeneous features, such as literally matching text, similarities in the
citations used, and similarity of mathematical content \cite{Meuschke14}. CitePlag is the first
system to implement such an integrated analysis and uses the VMEXT framework to visualize the
similarity of mathematical content.
\subsection{Demo Application}\label{sec:vmext:demo}
To showcase a possible integration of VMEXT's widgets into MIR and MKM applications, we developed a
Java application for input conversion and similarity computation. The demo provides a basic web
frontend available at: \url{http://vmext-demo.formulasearchengine.com} and offers two main features.

First, it converts \LaTeX\ input to parallel \MathML{}. The backend of the demo application offers
two alternative converters. The first converter employs \LaTeXML{}, whose configuration can be
customized via input fields included in the web frontend. The second converter passes the \LaTeX{}
input to the Mathoid system\footnote{\url{https://www.mediawiki.org/wiki/Mathoid}}
\autocite{disCicm14Mathoid}, which employs the speech rule
engine\footnote{\url{https://github.com/zorkow/speech-rule-engine}} \autocite{cervone2015towards} to
generate Presentation \MathML{} with {\bf CDATA} annotations. These annotations give hints on the
possible semantic meaning of expressions. Using a simple XSLT stylesheet, the demo application
converts this non-standard-conforming markup to standard parallel \MathML{} markup. The application
enables users to quickly run different \LaTeX\ to \MathML\ conversions and immediately examines the
effects on the conversion quality using the VMEXT visualizations described in Section
\ref{sec:vmext:ast} and Section \ref{sec:vmext:sim}.

Second, the demo application computes basic similarity measures for two expressions. The most basic
measure identifies identical nodes. A second measure uses the idea of taxonomic distance of
expressions proposed in \autocite{zhangY14}. Our implementation uses content dictionaries to model
the taxonomic distance and builds upon the content dictionary abstraction as introduced in
\autocite{disNtcir11Sim}. The system converts the CMML markup of the expression to Strict CMML to
guarantee that the XML encodings of all symbols explicitly state from which content dictionary the
symbols originate. All symbols originating from the same content dictionary, like plus and minus, or
sine and cosine, are considered similar. Symbols from different content dictionaries, e.g., plus and
cosine, are considered dissimilar. The objective of the similarity computation is to provide users
with test data to explore the visualization approaches, and not to be meaningful from an analytical
perspective.

\subsection{Obtaining VMEXT}\label{sec:vmext:use}
VMEXT is a free and open source JavaScript application. We host a ready-to-use instance of the tool
at: \url{http://vmext.formulasearchengine.com}. We also provide a REST API that exposes the image
export functionality and the internal representation of our visualization.

The demo application for converting and rendering \LaTeX\ markup (see Section \ref{sec:vmext:demo})
is available at: \url{http://vmext-demo.formulasearchengine.com}.

For development purposes,VMEXT is available as a Node.js package from:
\url{https://www.npmjs.com/package/vmext}. We actively maintain and enhance the tool; the latest
code is available from \url{https://github.com/ag-gipp/vmext}. Pull requests and bug reports are
highly welcome.

\section{Conclusion and Future Work}\label{sec:future-work}
In this paper, we present two tree-based visualization approaches for mathematical expressions. The
first approach simultaneously illustrates the presentation, structure, and semantics of individual
expressions. The second approach visualizes the structural and semantic similarity of two
expressions. Both approaches operate on parallel \MathML\ markup and incorporate key elements of
expression tree visualizations proposed in the MIR literature.

We implemented the two approaches as part of VMEXT, a system we provide free and open source for end
users and developers (see Section \ref{sec:vmext:use}). Additionally, we provide two web-based demo
applications. The first application\footnote{\url{http://vmext.formulasearchengine.com}} presents
the visualization widgets alone. The second
application\footnote{\url{http://vmext-demo.formulasearchengine.com}} demonstrates a possible
integration of the widgets in systems for mathematical knowledge management and mathematical
information retrieval.

In our future work, we plan to extend VMEXT's functionality beyond exclusively visualizing \MathML\
markup towards visually assisting markup creation and editing by humans. \MathML\ shows great
promise for enabling unprecedented access to mathematical knowledge. However, converting existing
mathematical knowledge to semantic markup formats will require some human interaction. The
complexity and verbosity of \MathML\ makes direct interaction with \MathML\ markup laborious and
time-consuming. We see visual editors as a possible solution to this problem. Enabling users to
create and manipulate mathematical notation and \MathML\ markup via visual support tools would be
valuable for increasing the digital accessibility of mathematical knowledge \cite{corneli17,
  disSigir16}. Another possible extension is the consideration of proof structures and the
visualization of the directed acyclic graphs, which might occur, if the \MathML\ \texttt{<share />}
element is used.
\paragraph*{Acknowledgments}
We thank Ludwig Goohsen and Stefan Kaufhold for their support in developing VMEXT. Furthermore, we
thank the Wikimedia Foundation for providing a server to run the VMEXT demo.
\vspace{-.5cm}
\printbibliography[keyword=primary]
\end{document}